\documentclass[conference]{IEEEtran}
\IEEEoverridecommandlockouts

\usepackage{cite}
\usepackage{amsmath,amssymb,amsfonts}
\usepackage{algorithmic}
\usepackage{graphicx}
\usepackage{textcomp}

\usepackage[table]{xcolor}
\newcommand{\act}{\cellcolor{lime!35}}  
\newcommand{\crit}{\cellcolor{yellow!45}\textbf{c}} 
\usepackage[hidelinks]{hyperref}
\def\BibTeX{{\rm B\kern-.05em{\sc i\kern-.025em b}\kern-.08em
    T\kern-.1667em\lower.7ex\hbox{E}\kern-.125emX}}
\DeclareUnicodeCharacter{202F}{\,}
\DeclareUnicodeCharacter{2011}{-}
\usepackage[hidelinks]{hyperref}
\begin{document}

\title{What is Cybersecurity in Space?}
\author{\IEEEauthorblockN{Charbel Mattar}
\IEEEauthorblockA{
\textit{LaRRIS, Faculty of Sciences} \\
\textit{Lebanese University} \\
Fanar, Lebanon \\
c.mattar@st.ul.edu.lb}
\and
\IEEEauthorblockN{Jacques Bou Abdo}
\IEEEauthorblockA{
\textit{School of Information Technology} \\
\textit{University of Cincinnati} \\
Cincinnati, OH, USA \\
bouabdjs@ucmail.uc.edu}
\and
\IEEEauthorblockN{Abdallah Makhoul}
\IEEEauthorblockA{
\textit{FEMTO-ST Institute, CNRS} \\
\textit{Université Marie et Louis Pasteur} \\
F-25200 Montbéliard, France \\
abdallah.makhoul@univ-fcomte.fr}

\and
\IEEEauthorblockN{Benoît Piranda}
\IEEEauthorblockA{
\textit{FEMTO-ST Institute, CNRS} \\
\textit{Université Marie et Louis Pasteur} \\
F-25200 Montbéliard, France \\
benoit.piranda@univ-fcomte.fr}
\and
\IEEEauthorblockN{Jacques Demerjian}
\IEEEauthorblockA{
\textit{LaRRIS, Faculty of Sciences, Lebanese University, Fanar, Lebanon} \\
\textit{Computer Science \& IT Department, Holy Spirit University of Kaslik} \\
Jounieh, Lebanon\\
jacques.demerjian@ul.edu.lb}
}

\maketitle
\noindent
\begingroup
\setlength{\fboxsep}{8pt}
\color{red}
\fbox{%
\begin{minipage}{0.97\linewidth}\footnotesize\color{black}
\textbf{IEEE Copyright Notice—}\\[0.2em]
© 2025 IEEE. Personal use of this material is permitted. Permission from IEEE must be obtained for all other uses, in any current or future media, including reprinting/republishing this material for advertising or promotional purposes, creating new collective works, for resale or redistribution to servers or lists, or reuse of any copyrighted component of this work in other works.\\[0.3em]
\textit{To appear in:} Proc.\ ACS/IEEE AICCSA 2025, Doha, Qatar, Oct 19–22, 2025.\quad \textit{DOI:} (to be assigned)
\end{minipage}%
}%
\endgroup
\par\medskip

\begin{abstract}
Satellites, drones, and 5G space links now support critical services such as air traffic, finance, and weather. Yet most were not built to resist modern cyber threats. Ground stations can be breached, GPS jammed, and supply chains compromised, while no shared list of vulnerabilities or safe testing range exists.

This paper maps eleven research gaps, including secure routing, onboard intrusion detection, recovery methods, trusted supply chains, post-quantum encryption, zero-trust architectures, and real-time impact monitoring. For each, we outline the challenge, why it matters, and a guiding research question. We also highlight an agentic (multi-agent) AI approach where small, task-specific agents share defense tasks onboard instead of one large model.

Finally, we propose a five-year roadmap: post-quantum and QKD flight trials, open cyber-ranges, clearer vulnerability sharing, and early multi-agent deployments. These steps move space cybersecurity from reactive patching toward proactive resilience.
\end{abstract}


\section{Introduction}

Space systems have become essential to everyday life, from GPS and communication to weather, agriculture, and defense \cite{Kavallieratos2023}. At the same time, Earth’s orbit is becoming crowded as companies launch hundreds of satellites monthly and cloud providers open antenna access to many users.

This expansion increases cyber risks. State-backed groups can exploit stolen credentials, inject malicious code, or jam signals \cite{Deloitte2024}. Space is also a contested military domain: China is developing a “kill web" while Russia has already disrupted satellites, notably the 2022 KA-SAT incident that disabled thousands of modems in Ukraine \cite{Saltzman2025}.

Protecting these assets is urgent. Priorities include closing security gaps, updating encryption, enabling onboard anomaly detection, and rehearsing recovery before larger constellations deploy.

Emerging architectures bring new risks. \emph{Modular robots}, satellite “building blocks" that dock or swap parts in orbit~\cite{Mattar2025}, add vulnerabilities through shared buses and ports, requiring strong identity checks and tamper-proof updates. AI-driven defense offers another path: instead of one large model, lightweight agents can each monitor specific tasks (radio, power, keys) and coordinate through a simple bus. This “agentic” design fits tight satellite limits and improves resilience \cite{Loevenich2024,Siew2023}.

Satellite development is also accelerating: CubeSats now reach orbit in under a year, and firms like Planet Labs have launched hundreds via agile methods \cite{Kong2022,PlanetLabs2023}. While boosting innovation, compressed timelines and small teams heighten cyber risk.

This paper identifies eleven research gaps, each formulated as a guiding research question. These gaps are analyzed in Section~\ref{sec:directions}, followed by conclusions and future directions in Section~\ref{sec:conclusion}.

\begin{figure}[ht]
  \centering
  \includegraphics[width=0.75\columnwidth]{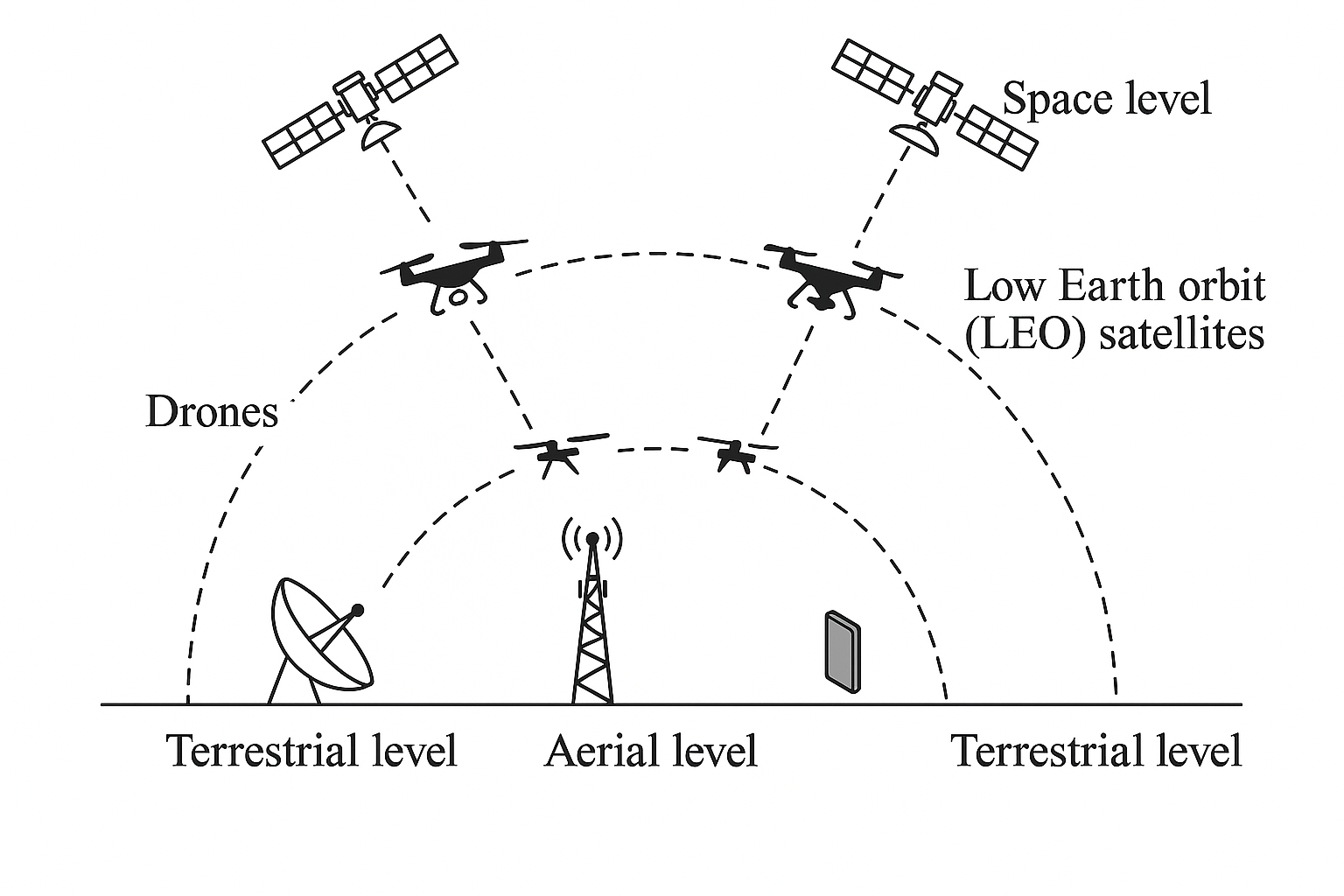}
  \caption{Multi-layer space, air, ground network architecture used in this
  study.  Geostationary and LEO satellites, drone relays, 5G/6G aerial
  towers, ground stations, and user equipment form an integrated
  communication mesh across space, aerial, and terrestrial domains.}
  \label{fig:layers}
\end{figure}

\section{Research Gaps and Directions in Space Cybersecurity}
\label{sec:directions}

Satellites, drones, ground stations, and 5G towers are now all connected in one large global network. To make this network secure from cyber threats, research is needed in the eleven gaps described below.

Each of the gaps involves one or more of the layers depicted in Fig.~\ref{fig:layers}'s space-air-ground architecture. For each gap, we describe the current situation, explain why it matters, and provide a guiding research question.  

\begin{figure}[ht]
  \centering
  \includegraphics[width=0.65\linewidth]{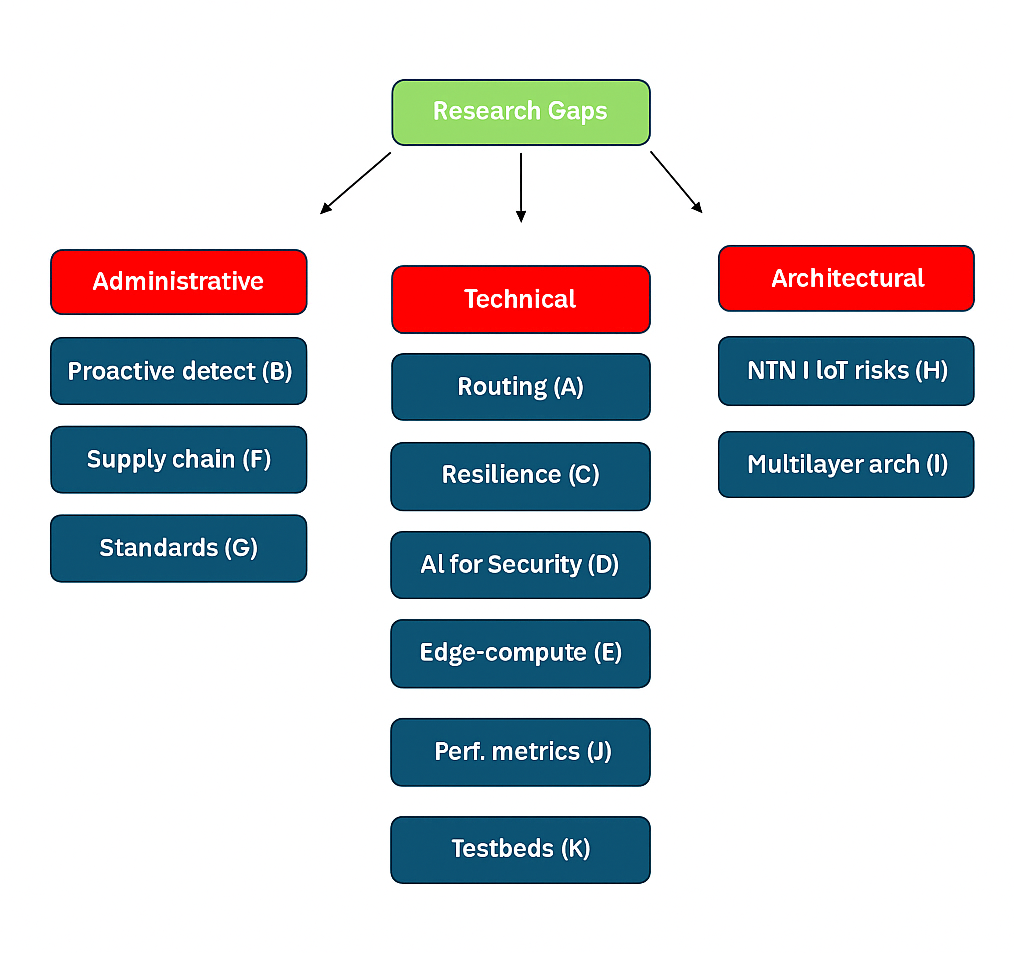}
  \caption{Classification of the eleven research gaps into three categories: Administrative, Technical, and Architectural. Letters (A-K) correspond to the research gap subsections in Section~\ref{sec:directions}.}

  \label{fig:research_gaps_tree}
\end{figure}

Fig.~\ref{fig:research_gaps_tree} illustrates the classification of the eleven research gaps into Administrative, Technical, and Architectural categories.

\subsection{Routing and Communication Security}

Satellites send data to each other using laser or RF cross-links that change frequently. Many of these links still update routing information without strong authentication \cite{Bhattacharjee2024,Jiang2015}. An attacker could send fake routing updates or block one link, causing data to be lost, like in BGP attacks on the internet.

As these satellite constellations get bigger and more dynamic, routing protocols need to adapt quickly while staying secure and efficient. In space, size and energy limits make this harder. Traditional methods like IPsec are too heavy for small satellites, and post-quantum cryptography is still being developed for fast-changing space links \cite{Tedeschi2022}.

A recent review~\cite{Mattar2025b} compared over 20 models that simulate how information spreads in IoT, drone, and space networks. Tools like NS-3 and OMNeT++ are helpful, but they don’t support space-specific problems like radiation, energy-aware routing, or attacks.

Improving these simulators and adding space-specific features can help researchers test new routing ideas. The review also pointed to the need for hybrid models that combine fixed routing with smart AI backups for more secure data delivery in space.

\textbf{Research question:} \emph{Which slim, quantum-ready routing scheme can keep packets safe even when links move or drop without warning, and how can we simulate its performance under space-specific failures?}

\subsection{Proactive Cybersecurity Measures}
Many small satellites still send raw logs to ground control and rely on human operators to detect intrusions, which can take hours \cite{Diro2024}. In that time, an attacker might hide their tracks, damage systems, or change the satellite's orientation.

Recent research on space systems, IoT, and edge computing shows that small, radiation-tolerant computers and lightweight AI chips are now being tested for use in CubeSats \cite{Chien2019,Kua2021}. This makes it possible to run simple machine learning models onboard that can detect unusual commands, power spikes, or strange radio signals right away.

However, there is still no clear method for training, updating, and validating these models after the satellite is launched.

\textbf{Research question:} \emph{What kind of ultra-low-power AI system can run directly on the satellite, detect attacks in seconds, and activate safe mode before the ground team even sees the alert?}

\subsection{Low Resilience and Recovery After Attacks}

The KA‑SAT incident forced thousands of terminals to be manually re-flashed \cite{Salim2025}. Unlike servers on Earth, satellites cannot be fixed by plugging in a USB stick. Once in space, they must handle recovery and coordination on their own.

Techniques like mesh rerouting, automatic key changes, and direct tunnels between satellites could help others take over when one fails. But we still don’t have much real-world data showing how well these methods work in space conditions. A recent SmartSat CRC study identified security layers—like radio links, hardware, ground stations, and operations—and stressed that automatic switching and redundancy are key for resilience \cite{Plotnek2022}.

Some researchers have proposed using blockchain-style consensus between satellites, allowing them to agree on which ones need fixing or isolating \cite{Ling2020}. Delay-Tolerant Networking (DTN), described in RFC 4838, can store and forward data when the network is slow or broken. However, it hasn’t been widely tested in orbit \cite{Cerf2007}. New ideas like POAST combine DTN with energy-saving, fault-tolerant consensus protocols \cite{Patil2024}, and trust systems can help filter out compromised nodes during emergencies \cite{Asuquo2018}.

\textbf{Research question:} \emph{What combination of repair strategies will help a satellite network recover quickly from a cyberattack? How can we coordinate these systems to stay resilient without adding too much complexity or weight?}

\subsection{Using AI and Agents for Security}

Tests in the lab show that AI can cut the time it takes to spot a cyber-attack by about 40\%, but real satellites have major limits: Satellites often lack enough labeled training data, and their limited communication bandwidth and power make it hard to use large AI models that need heavy GPU hardware \cite{Diro2024}.
Federated learning can help, where each satellite trains on its own data and sends back only tiny updates \cite{Hajj2023}. It has not been tested if federated learning performs under space constraints such as radiation hits, power cuts, or week-long gaps in contact.  

\textbf{Research question:} \emph{How can solid AI models be trained or updated when bandwidth is limited and good training examples are rare?}

\paragraph*{Agentic Solution}
To address this research question, an agentic (multi-agent) approach could replace one large AI model with many tiny AI agents, each focused on a single task. This setup is lighter, easier to swap, and more fault-tolerant.

\begin{itemize}
  \item \textbf{RF Guard} - listens for strange radio tones;  
  \item \textbf{Power Watch} - flags odd battery drain;  
  \item \textbf{Key Minder} - rolls new crypto keys on a timer;  
  \item \textbf{Update Checker} - makes sure code upgrades are signed.  
\end{itemize}

All agents communicate over a simple “message bus,” and a tiny traffic-cop (an \emph{agent-policy engine}) decides who can send which command.  
If an agent crashes, we swap in a new one without touching the rest, perfect for space hardware where reboots are risky.  

\begin{figure}[t]
  \centering
 \includegraphics[width=0.45\columnwidth]{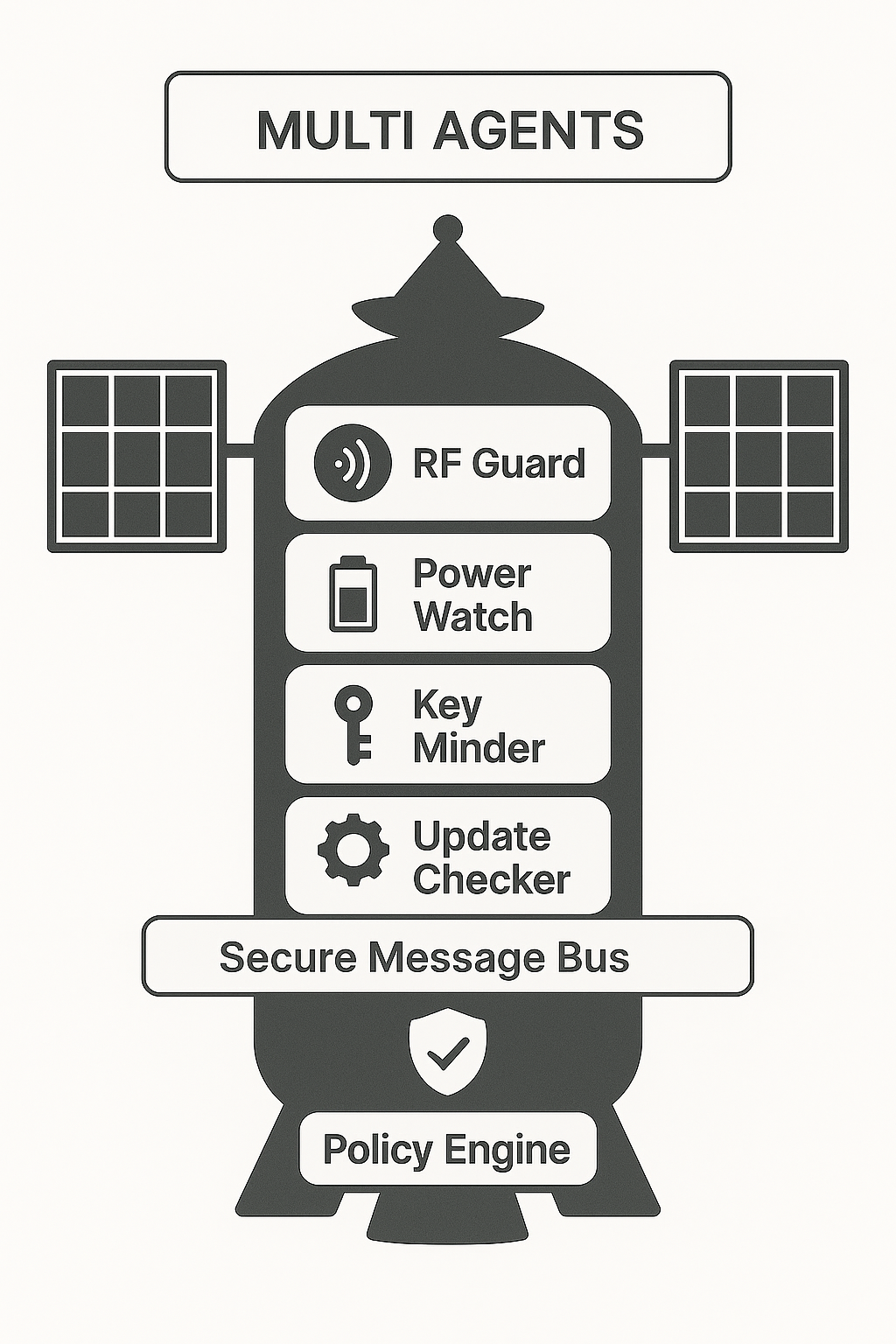}
  \caption{Lightweight multi-agent layout.  
           Four task-specific agents share a secure message bus policed by a tiny policy engine.}
  \label{fig:onboard_agents}
\end{figure}

Fig.~\ref{fig:onboard_agents} shows a simple layout of how four task-specific AI agents can share a secure message bus onboard a satellite, coordinated by a lightweight policy engine.

Field trials are starting to arrive: a hybrid AI defense agent that mixes rules, reinforcement learning, and Large-Language Model (LLM) is already protecting a test network in the lab \cite{Loevenich2024}.  
A separate study shows that dozens of reinforcement-learning agents can share tasks across hundreds of space objects without choking the link budget \cite{Siew2023}.  

\paragraph*{Open problems for agent swarms}
\begin{enumerate}
  \item \textbf{Trust:}  How does RF Guard know Key Minder (the tiny Ai agents) has not been hacked? Lightweight identity checks (hash-chains or a micro-ledger) may help.
  \item \textbf{Updates:}  Agents must learn new tricks after launch. Distilling a bigger ground model down to a small on-board agent, while links are slow, needs smart compression.
  \item \textbf{Coordination:} During disruptions, such as a solar flare that disables many sensors, the system must quickly determine which agent takes primary control and which agents switch to standby. Formal coordination methods, for example a three-agent majority vote, can prevent control deadlocks.
\end{enumerate}

\textbf{Second research question:} \emph{What light but safe handshake lets many on-board agents work together without giving a hacker one easy target to hijack?}

\subsection{Edge and On-Board Computing}
Recent studies on space IoT show that many CubeSat-sized boards now come with built-in GPUs or vision processors that can handle trillions of operations per second. But in most missions, these chips are barely used. The satellites usually just collect data and send it back to Earth as telemetry \cite{Kua2021,Chien2019}.

This unused computing power could be helpful. It might verify cryptographic hashes, detect strange radio signals, or run a basic zero-trust security tool directly onboard the satellite.

However, there are some real limits. Running extra software could drain the battery, produce too much heat, or interfere with the main mission. Also, any added code must go through strict safety checks before it’s used in space.

\textbf{Research question:} \emph{How can we safely use extra onboard computing power to improve cybersecurity, without affecting the satellite’s main tasks?}

\subsection{Quantum Computing}
Quantum computing brings serious long-term risks to satellite networks. Most space data today is protected by RSA and ECC encryption, but these could be broken once powerful quantum computers become available. That means attackers might record encrypted data now and decrypt it years later when the technology improves, a method known as “harvest now, decrypt later.”

Post-quantum (PQ) algorithms are being developed to resist quantum attacks, and they can run on regular CPUs. However, they often need more memory, larger keys, and longer handshake times. That’s a challenge for spacecraft, which must meet strict size, weight, and power (SWaP) limits \cite{Kumar2022}. It’s unlikely that one single PQ standard will fit all mission types.

Meanwhile, some real progress has been made. A 4,600 km Quantum Key Distribution (QKD) network has already linked satellites to fiber networks on the ground \cite{Chen2021}. QKD offers very strong security but relies on precise optical alignment, which is difficult due to satellite movement and atmospheric interference \cite{Li2019}.

Satellites launched in 2024 must still be secure in 2040. That means we need a combined roadmap using both PQ cryptography and QKD hardware, with options to switch depending on the mission.

\textbf{Research question:}
\emph{What is the best combination of PQ cryptography and QKD technology to keep satellite networks secure across their full lifespan?}

\subsection{Global Cybersecurity Rules or Standards}
Some countries require breach disclosures within 24 hours, while others have no rule at all \cite{RacioneroGarcia2024}. Without a shared “Space-CVE” list (like the Common Vulnerabilities and Exposures list used on Earth), fixing the same security issue may take longer and cost more across different organizations \cite{Falco2019}.

Recent progress in cyber defense agents shows that using shared frameworks, like MITRE~ATT\&CK, OpenC2, and structured knowledge graphs, helps different systems work together \cite{Loevenich2024}. For example, NATO uses these shared tools to help agents explain and respond to attacks across multiple networks.

Having standard names for vulnerabilities, common response plans, and coordinated testing could reduce confusion and speed up defense. The hard part is that many countries still don’t fully trust each other.

\textbf{Research question:}
\emph{What is the smallest, politically realistic cybersecurity standard that all space operators can agree on for naming, sharing, and fixing security problems?}

\subsection{New Tech Connections Add Risks (IoT, 5G/6G..)}
5G Non‑Terrestrial Network (NTN) pilots let a standard phone ping a LEO
satellite through control channels \cite{Zhou2024NTN}.  Drones acting as
edge relays widen the bridge between Earth malware and orbital routers.
\\
Zero‑trust architecture, authenticate every hop, encrypt everywhere, works in data centers; doing it across moving satellites, drones, and towers is uncharted.

\textbf{Research question:} \emph{How can zero‑trust ideas stretch seamlessly from a phone on Earth, through 5G slices and drone relays, to an in‑orbit router?}

\subsection{Multilayer Security Strategies}
Strong crypto often shields the ground link, yet payload buses or drone
hops may still use legacy protocols \cite{Wu2023}.  Attackers go for the weakest link, so policy must span ground, air, and space. 

Designers need a reference stack, keys, roles, update paths, that treats the whole chain as one fabric.

\textbf{Research question:} \emph{What end‑to‑end architecture enforces
the same key‑management and access rules across satellites, drones, and
ground stations?}

\subsection{Performance During Attacks}
Deliberate jamming or spoofing forces packets to detour in a LEO
constellation and this can push end-to-end delay from the usual 20 - 40 ms up to 200 ms or more \cite{Wang2024}. Such jumps freeze tele-operation and
other tight-loop applications.\\
Two recent surveys point out that, although threat reports are
plentiful, very few missions track \emph{live} service-impact numbers:
operators still lack a quick dial that says “this attack just cost us
30 \% quality of service" \cite{Salim2025,Kavallieratos2023}.  Without a
small, shared KPI set, teams cannot compare defenses or decide when to
switch to safe-mode.

\textbf{Research question:} \emph{Which handful of easy-to-read
metrics, such as latency, burst packet-loss, or a user-experience score, best
captures the real cost of a cyber or RF attack, which can be reported in
real time?}

\subsection{Tools to Measure and Track Security}
Although many papers discuss satellite cyberattacks, there are few open testbeds where researchers can run realistic “live-fire" drills without risk \cite{Kavallieratos2023,Salim2025}. Existing simulators such as NS-3 and OMNeT++ support mobility and delay-tolerant routing but lack features like radiation effects, GPS spoofing, or cross-layer attacks \cite{Mattar2025b}.

There is also no shared way to measure resilience. Benchmarks for recovery time, detection speed, or service continuity under denial-of-service are missing, making solutions hard to compare or validate.

A modular, open-access cyber range with hardware-in-the-loop, simulated telemetry, and customizable attack tools could provide repeatable testing.

\textbf{Research question:} \emph{What kind of modular, open cyber range can let researchers test satellite cyberattacks, compare recovery techniques, and agree on shared metrics?}

\begin{table}[ht]
\centering
\caption{Layer-wise relevance of the eleven research gaps.  
T = primary technical focus; \crit\ = layer where consequences are most critical; “-” = not a main driver.}
\label{tab:gaps_layers}
\setlength{\tabcolsep}{4pt}
\renewcommand{\arraystretch}{1.25}
\small
\begin{tabular}{l|ccccc}
\hline
\textbf{Gap} & \textbf{Terr.} & \textbf{Aerial} & \textbf{LEO} & \textbf{GEO} & \textbf{Deep-Space}\\\hline
Routing (A)                 & - & \act & \act & \act & - \\
Proactive detect (B)        & - & \act & \act & \act & \act \\
Resilience (C)              & - & -    & \crit& \crit& \act \\
AI for Security (D)         & - & \act & \act & \act & \act \\
Edge-compute (E)            & - & -    & \act & -    & - \\
Supply chain (F)            & \act & \act & \act & \act & \act \\
Standards (G)               & \act & \act & \act & \act & \act \\
NTN / IoT risks (H)         & \act & \act & \act & -    & - \\
Multilayer arch (I)         & \act & \act & \act & \act & \act \\
Perf. metrics (J)           & - & \act & \act & \crit& - \\
Testbeds (K)                & \crit& \act & \act & \act & - \\\hline
\end{tabular}
\end{table}

Table~\ref{tab:gaps_layers} divides the eleven research gaps into a single cross-layer map.  A green cell marks layers where active technical effort is already under way,
whereas a yellow cell
(\crit) flags the layer whose mission would suffer the most should that
gap remain unaddressed.  Reading \emph{down} a column shows which gaps
cluster in a given domain (terrestrial segments are dominated by
supply-chain and standardization issues), while reading \emph{across} a
row reveals how broadly a single gap spans the architecture.
Gaps that receive both an active-research mark (green) and a critical-impact mark (yellow ‘c’) in the same layer, such as
\emph{Resilience} and \emph{Performance metrics} in LEO/GEO, represent high investment targets for the next five years, whereas those
with scattered \act\ cells can be scheduled for longer-term research
programmes.  This cross-layer view therefore complements the
gap-by-gap discussion in Section~\ref{sec:directions} by highlighting
where defenses must converge first.
\section{Conclusion}
\label{sec:conclusion}

Space cybersecurity is evolving rapidly, with challenges spanning three areas: administrative (policy, disclosure, governance), technical (algorithms, encryption, onboard hardware), and architectural (end-to-end protection from space to ground). These issues affect not only engineering but also international cooperation. Awareness is growing, and first efforts to address them are underway.

Key lessons emerge from this study. Space hardware must remain secure for decades, requiring lightweight yet quantum-resistant methods that anticipate future threats. Satellites can no longer depend solely on ground teams; anomaly detection must occur directly onboard so a system can react within seconds rather than hours. At the same time, shared rules and open test ranges remain absent, which slows the adoption of fixes and makes it harder to validate solutions across different operators.

Looking ahead, a five-year roadmap can guide action. The first step is to design inter-layer blueprints that treat satellites, drones, and ground nodes as one fabric, ensuring that keys, patches, and incident-response signals propagate seamlessly. A second priority is the deployment of lightweight AI agents directly on spacecraft and edge nodes to detect attacks, protect encryption keys, and support rapid recovery. Third, cyber-resilience metrics should become a standard element of constellation design reviews so that missions shift from patching after launch to secure-by-design principles. A fourth step is to demonstrate post-quantum cryptography and quantum key distribution in space-to-ground links, proving that future-proof security can be achieved under strict size, weight, and power limits. Finally, a politically neutral framework for naming and disclosing vulnerabilities must be established so that mitigation efforts are coordinated across orbital, aerial, and terrestrial segments.

These measures can move the sector from reactive patching toward proactive resilience, ensuring that humanity’s newest layer of infrastructure does not become its largest attack surface.

\end{document}